\documentclass[aip,jcp,amsmath,amssymb,longbibliography,floatfix,reprint]{revtex4-2}

\usepackage{graphicx}
\usepackage{dcolumn}
\usepackage{bm}

\usepackage[utf8]{inputenc}
\usepackage[T1]{fontenc}
\usepackage{etoolbox}
\usepackage{siunitx}
\usepackage{chemformula}
\usepackage{braket}
\usepackage[hidelinks]{hyperref}
\usepackage{xcolor}

\newcommand{\oper}[1]{\hat{#1}}

\makeatletter
\def\@email#1#2{%
 \endgroup
 \patchcmd{\titleblock@produce}
  {\frontmatter@RRAPformat}
  {\frontmatter@RRAPformat{\produce@RRAP{*#1\href{mailto:#2}{#2}}}\frontmatter@RRAPformat}
  {}{}
}%
\makeatother
\begin{document}

\title{Quantum critical behavior in chains of hindered dipolar planar rotors}

\author{Wenxue Zhang}
\affiliation{Department of Chemistry, University of Waterloo, Waterloo, Ontario, N2L 3G1, Canada}
\author{Muhammad Shaeer Moeed}
\affiliation{Department of Physics, University of Waterloo, Waterloo, Ontario, N2L 3G1, Canada}
\affiliation{Institute For Quantum Computing, University of Waterloo, Waterloo, Ontario, N2L 3G1, Canada}
\affiliation{Perimeter Institute for Theoretical Physics, Waterloo, Ontario N2L 2Y5, Canada}
\author{Estevao De Oliveira}
\affiliation{Department of Physics, University of Waterloo, Waterloo, Ontario, N2L 3G1, Canada}
\author{Hui Li}
\affiliation{
State Key Laboratory of Supramolecular Structure and Materials, Institute of Theoretical Chemistry, College of
Chemistry, Jilin University, 2519 Jiefang Road, Changchun 130023, P. R. China}
\affiliation{Department of Chemistry, University of Waterloo, Waterloo, Ontario, N2L 3G1, Canada}
\author{Pierre-Nicholas Roy}
\affiliation{Department of Chemistry, University of Waterloo, Waterloo, Ontario, N2L 3G1, Canada}
\affiliation{Institute For Quantum Computing, University of Waterloo, Waterloo, Ontario, N2L 3G1, Canada}
\affiliation{Perimeter Institute for Theoretical Physics, Waterloo, Ontario N2L 2Y5, Canada}
\email{pnroy@uwaterloo.ca}

\date{\today}

\begin{abstract}
We study the ground-state properties of linear chains of dipolar planar rotors hindered by a six-fold on-site potential, a model motivated by water molecules confined in the hexagonal cavities of beryl. Using density matrix renormalization group (DMRG) calculations, we locate the quantum phase transition between the disordered and ferroelectrically ordered phases using the von Neumann entanglement entropy and the Binder ratio of the polarization. A sweep of the six-fold pinning strength shows that increasing hindrance shifts the critical dipolar coupling $g_c$ to smaller values. These results suggest that crystal-field hindrance can promote, rather than suppress, dipolar ordering. This work has implications for ferroelectricity and quantum-device tuning in confined molecular rotors.
\end{abstract}

\maketitle
\section{Introduction}
Molecules confined in cavities of crystal structures such as beryl and cordierite can host a number of interesting quantum phases,\cite{kolesnikov2016quantum,finkelstein2017quantum,klivc2023orientational,belyanchikov2022single,gorshunov2013quantum,zhukova2014vibrational,gorshunov2016incipient,zhukova2019quantum,belyanchikov2020dielectric} owing to the guest molecules' rotational degrees of freedom. For instance, previous work has suggested that confining water molecules in beryl suppresses the hydrogen bonding typically observed between water molecules. This suppression is caused by the small cavity size of the beryl structure.\cite{anovitz2013anisotropic,Gorshunov02112014,kolesnikov2016quantum,dressel2018quantum,kolesov2000orientation,kolesov2008vibrational}
This can allow the lattice of pinned water molecules to exhibit ferroelectric order, because of dipolar interactions between different (occupied) sites in the crystal.\cite{farimani2013rotational,gorshunov2016incipient,zhao2014ferroelectric}

The orientational ordering phenomena of confined molecules are well-described by a lattice of rotors with dipolar interactions.\cite{abolins2011ground,serwatka2023quantum,vb2025path} In particular, for pinned molecules in fullerenes, rotor lattices have proven to be a quantitatively accurate platform for examining the phase diagram in $1d$ chains.\cite{serwatka2022ferroelectric,serwatka2022ground,serwatka2023qpt,serwatka2024quantum,serwatka2024ground} Several computational and theoretical techniques have been developed in the past to study planar rotor assemblies.\cite{serwatka2023qpt,serwatka2023quantum,serwatka2024quantum,zhang2025path}
Relevant examples include the Density Matrix Renormalization Group (DMRG)\cite{white1992density,schollwock2011density} as well as optimized path integral simulation techniques such as Gibbs sampling in a discrete variable representation (DVR).\cite{xiao2007using,zhang2025path} DMRG was used to study the critical point of planar dipolar rotor chains with a $\mathbb{Z}_2$ symmetry.
It was shown that these exhibit a $(1+1)d$ Ising universality class transition at a critical coupling strength $g_c\approx 0.5$.\cite{serwatka2024quantum}

In crystals such as beryl, the confining cavity has a hexagonal structure.\cite{klivc2023orientational} Moreover, these cavities are typically only large enough to accommodate a single water molecule each. Recent Inelastic Neutron Scattering (INS) experiments suggest that due to the hexagonal structure of the confining cavity,\cite{kolesnikov2016quantum,anovitz2013anisotropic} the crystal structure cannot be entirely neglected. 
Additionally, the distance between the confined molecules and the cage is relatively small.
The cavity geometry constrains the dipole along the six edges of the hexagonal cage. This crystal-induced anisotropy can be effectively modeled using a six-fold on-site local potential. This is consistent with the close alignment between the experimental and the six-fold potential model low-energy spectra.\cite{finkelstein2017quantum,gorshunov2016incipient,klivc2023orientational,belyanchikov2022single} While DMRG is ideally suited to the effectively one-dimensional chains considered here, we note that extensions to higher-dimensional lattices will require alternative approaches such as path integral Monte Carlo.\cite{ceperley1995path,marx1999path,abolins2011ground,abolins2013erratum,abolins2018quantum,sahoo2021path,sahoo2023effect,zhang2025path,Moeed2025,vb2025path}

\begin{figure}[htb]
    \centering
    \includegraphics[width=\linewidth]{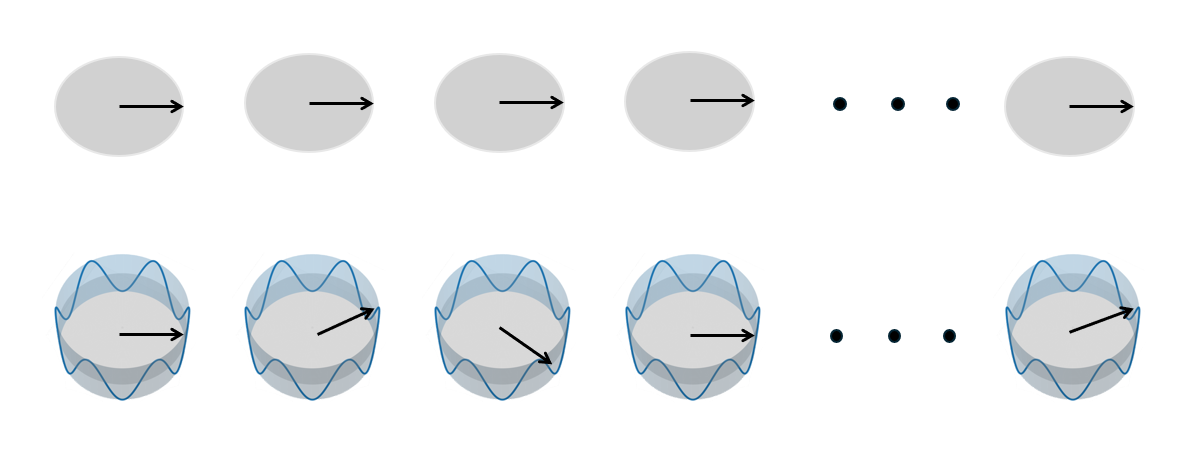}
    \caption{Schematic of a coplanar chain of dipolar planar rotors, each subject to a six-fold hindering potential.}\label{paper2:fig:rotor_chain_with_potential}
\end{figure}

In the present work, we perform large-scale DMRG simulations to examine the effect of this six-fold potential on the phase diagram of a coplanar chain of dipolar rotors (see Fig.~\ref{paper2:fig:rotor_chain_with_potential}). We find that in the $\mathbb{Z}_2$ chain, the strength of the constraining six-fold potential shifts the critical point in the dipolar coupling strength $g$. Increasing the strength of the six-fold potential shifts the onset of ferroelectric order to smaller values of $g$. This implies that a larger class of confined molecular species than previously considered has the potential to exhibit spontaneous alignment of dipole moments. To study this quantitatively, we use the Binder ratio\cite{binder1981finite,binder1981critical,serwatka2023qpt} of the polarization, as well as the entanglement entropy.\cite{calabrese2004entanglement,amico2008entanglement,eisert2010area} These two measures exhibit a sharp transition from the disordered to the ordered phase and are key quantities for the determination of critical points in $(1+1)d$ Ising universality class transitions.\cite{sachdev2011quantum,iouchtchenko2018ground,serwatka2023qpt,serwatka2024quantum}

The remainder of this paper is organized as follows: in Sec.~\ref{paper2:theory}, we discuss the exact form of the Hamiltonian and the behavior of the ground state in various limits. In Sec.~\ref{paper2:results}, we quantitatively characterize the effect of the six-fold potential on the critical point using our DMRG calculations. Finally, in Sec.~\ref{paper2:conclusions}, we conclude and discuss future research directions.

\section{Theory\label{paper2:theory}}

\subsection{System Hamiltonian}

We consider a one-dimensional ($1d$) chain of $N$ planar rotors with nearest-neighbor dipolar interactions. Each rotor is hindered by a six-fold (hexagonal) on-site potential. This potential models the anisotropy induced by the background crystal structure. The system Hamiltonian is
\begin{align}
    H &= \sum_{i=1}^{N}\left( \frac{B}{\hbar^2} \oper{L}_{z,i}^2 -\frac{A'}{2}\big(\cos(6\phi_i)-1\big)\right) \nonumber\\
    &+\frac{|\mu|^2}{4\pi\epsilon_0 R^3}\sum^{N-1}_{i=1} \hat{V}_{i,i+1} ,
    \label{paper2:eq:H}
\end{align}
where $\oper{L}_{z,i} = -i\hbar \partial_{\phi_i}$ is the planar rotor angular momentum operator and $\phi_i$ is the angle that parametrizes the orientation of rotor $i$. The parameter $B$ is the rotational constant of the rotor, $A'$ is the barrier height of the six-fold hindering potential, $\mu$ is the dipole moment of a rotor, and $R$ is the inter-rotor distance.

The nearest-neighbor dipolar interaction, $\hat{V}_{i, i+1}$, is defined as
\begin{equation}
    \hat{V}_{i,i+1} = \left[\hat{y}_i \hat{y}_{i+1} - 2\hat{x}_i \hat{x}_{i+1}\right],
\end{equation}
where the constrained Cartesian position operators $(\hat{x}_i, \hat{y}_i)$ of each rotor are conveniently expressed in polar coordinates as $(\cos(\phi_i), \sin(\phi_i))$. 
With this sign convention, head-to-tail alignment of the dipoles along the chain axis is energetically favored. This alignment determines the structure of the ordered phase discussed below. This representation of the problem is depicted in Fig.~\ref{paper2:fig:rotor_chain_with_potential}, which shows a chain of rotors, each with its own six-fold localization potential.

Note that the $1/R^3$ dipolar interaction is truncated at nearest neighbors. The neglected next-nearest-neighbor coupling is suppressed by a factor of $2^3 = 8$. We therefore expect only a small quantitative error. 
Also note that the six-fold potential of Eq.~\eqref{paper2:eq:H} places two of its six minima along the chain axis ($\phi = 0, \pi$). These are the orientations favored by the dipolar interaction. This alignment maximizes the cooperation between hindrance and dipolar order. A relative misalignment between the crystal-field minima and the chain axis would weaken this cooperation and may lead to different order. A quantitative study of this effect will be the subject of future work.

It is convenient to work in reduced units by dividing Eq.~\eqref{paper2:eq:H} by $B$. Since the six-fold potential is a single-body operator, it can be absorbed into the monomer Hamiltonian
\begin{equation}
    \hat{h}^{1\mathrm{B}}_i = -\frac{\partial^2}{\partial \phi^2_i} \underbrace{- \frac{A}{2} \big(\cos(6\phi_i)-1\big)}_{\hat{V}_{6,i}},
    \label{paper2:eq:h1B}
\end{equation}
where $A=A'/B$ denotes the reduced barrier height. Defining the dimensionless dipolar coupling strength
\begin{equation}
g=\frac{|\mu|^2}{B~4\pi\epsilon_0 R^3},
\end{equation}
the reduced Hamiltonian takes the compact form
\begin{equation}
    \frac{\hat{H}}{B} = \sum_{i=1}^{N} \hat{h}^{1\mathrm{B}}_i + g \sum_{i=1}^{N-1} \hat{V}_{i,i+1}. \label{paper2:eq:HB}
\end{equation}
Note that no separate definition of the six-fold potential is required, as the hindering term is contained entirely in $\hat{h}^{1\mathrm{B}}_i$ of Eq.~\eqref{paper2:eq:h1B}.

For our DMRG calculations, we use the momentum basis, which diagonalizes the angular momentum operator as follows 
\begin{equation}
    \hat{L}_{z,i}^2 = \sum_{m=-\infty}^{\infty} m^2 |m\rangle \langle m|.
\end{equation}
Previous work has examined ground-state properties of free planar rotor chains with a dipole interaction.\cite{serwatka2024quantum} There, a symmetric momentum cutoff $l = m_{\max} = -m_{\min} = 4$ was shown to yield quantitatively accurate results.
For the chain of hindered rotors with the same dipole interaction, we use a similar approach. However, since the spectrum behaves differently in the presence of the six-fold potential, we anticipate that the value of $l$ needed for convergence will be different. Based on the monomer convergence analysis of Sec.~\ref{paper2:results}, we adopt $l = m_{\mathrm{max}} = 16$ in this work. 

In this finite momentum representation, the dipole interaction takes the form 
\begin{gather}
    \hat{V}_{i,i+1} = -\frac{1}{4} \sum_{m_i,m_{i+1}} |m_{i} + 1, m_{i+1} - 1 \rangle \langle m_{i}, m_{i+1} | \nonumber \\ - \frac{3}{4} \sum_{m_i, m_{i+1}} |m_{i} + 1, m_{i+1} + 1 \rangle \langle m_{i}, m_{i+1} | + h.c., 
\end{gather}
where the sums over $m_{i}$ and $m_{i+1}$ go from $-l$ to $l$. In the same basis, the total six-fold hindering potential, defined through $\hat{V}_{6,i}$ of Eq.~\eqref{paper2:eq:h1B}, is represented as
\begin{equation}
    \sum_{i=1}^{N} \hat{V}_{6,i} = -\frac{A}{4} \sum_{i=1}^{N} \left( \sum_{m_{i}=-l}^{l} |m_i + 6 \rangle \langle m_i| + h.c. - 2  \right),
\end{equation}
while the kinetic term is diagonal with eigenvalues $m_i^2$. Under the momentum cutoff, matrix elements with $|m_i + 6| > l$ are discarded.

\subsection{Estimators}

In this work, we are interested in the ground-state phase diagram of the system defined above. For the dipolar chain of free rotors ($A = 0$), previous work has shown that there is a $(1+1)d$ Ising universality class Quantum Phase Transition (QPT) at $g \approx 0.5$.\cite{iouchtchenko2018ground,serwatka2023qpt,vb2025path} This transition separates the disordered ($g < 0.5$) and ordered ($g > 0.5$) phases, in the thermodynamic limit ($N \rightarrow \infty$). 
Note that the dipolar interaction explicitly breaks the $U(1)$ symmetry of the free rotor system down to $\mathbb{Z}_2$. As $g$ increases, this QPT results as a consequence of the spontaneous breaking of this $\mathbb{Z}_2$ symmetry.

To study the phase diagram and the behavior of this transition in the presence of the six-fold potential, we use the Binder ratio and the entanglement entropy. The former is typically used in many-body physics for accurately estimating the critical point.\cite{binder1981finite,binder1981critical,serwatka2023qpt} It is defined as follows: 
\begin{equation}
    U_x=1-\frac{\braket{M_x^4}}{3\braket{M_x^2}^2}, 
\end{equation}
where $M_x = \sum_{i=1}^{N} \cos(\phi_i)/N$ is the polarization of the rotors along the chain. In the ordered phase, the Binder ratio approaches $U_x = 2/3$.
It serves as a sharp diagnostic of the QPT, for which the polarization $M_x$ is the order parameter. 

Note that deep in the disordered phase, as $N \rightarrow \infty$, we expect the polarization to be normally distributed since the ground state is well approximated by the limit: 
\begin{equation}
    |\psi_0 \rangle = \bigotimes_{i=1}^{N}|E_0^i\rangle. \label{paper2:gs_disorder}
\end{equation}
Here, $|E_0^i \rangle$ is the ground state of the $i$th rotor.
This can be seen by taking the limit $g \rightarrow 0$. That limit yields a lattice of non-interacting rotors.
In the absence of the six-fold potential, this is simply the angular momentum eigenstate $|m=0\rangle$. Since the rotors in this limit are decoupled, $M_x$ is the average of $N$ independent and identically distributed random variables. Therefore, $M_x$ tends to a normal distribution in the thermodynamic limit. The Binder ratio is $0$ by definition in that case. 

Conversely, in the presence of $\mathbb{Z}_2$ order, the ground state for a finite chain in the limit of $g \rightarrow \infty$ is: 
\begin{equation}
    |\psi_0 \rangle = \frac{1}{\sqrt{2}} \left( \bigotimes_{i=1}^{N} |\phi_i = 0 \rangle + \bigotimes_{i=1}^{N} |\phi_i = \pi \rangle \right). \label{paper2:gs_order}
\end{equation}
To see this, note that the orientations associated with each state in the superposition above correspond to the minima of the dipolar interaction term. That term dominates the Hamiltonian in the limit $g \rightarrow \infty$.
It is then clear that $\langle M_x^4 \rangle = \langle M_x^2 \rangle = 1$, which corresponds to $U_x = 2/3$. 
It is instructive to note that while our discussion here refers to a chain with finite $N$, the QPT only exists in the limit of $N \rightarrow \infty$. 
Moreover, the spontaneous development of order implies that one of the two states in the superposition in Eq. \eqref{paper2:gs_order} then represents the ground state. This yields a non-zero order parameter (polarization). However, the conclusions about the Binder ratio remain the same in this case since it is an even function of the polarization. This is also why it is an effective probe of the QPT and the ordered phase in finite size systems with large $N$.

Another useful measure of critical phenomena in this system is the bipartite von Neumann entanglement entropy defined as: 
\begin{equation}
    S_{\mathrm{vN}} = -\text{Tr}_{A}(\rho_{A} \ln (\rho_{A})), \ \ \ \rho_{A} = \text{Tr}_B(\rho). 
\end{equation}
Here, $\rho_A$ is the reduced density matrix corresponding to subsystem $A$, and $\text{Tr}_A$, $\text{Tr}_B$ represent the partial trace over subsystems $A$ and $B$, respectively. At $T=0$, the density matrix of the entire system is simply determined by the ground state $|\psi_0\rangle$ as: $\rho = |\psi_0\rangle \langle \psi_0|$. For our analysis, we will set subsystems $A$ and $B$ to be the first and second sets of consecutive $N/2$ sites in the $N$ rotor chain, respectively. In the disordered phase, since the ground state (see Eq. \eqref{paper2:gs_disorder}) is separable, the entanglement entropy is $0$. In the ordered phase, for finite $N$, note that: 
\begin{equation}
    \rho_A = \frac{1}{2} \left( \bigotimes_{i=1}^{N/2} |0_i\rangle \langle 0_i| + \bigotimes_{i=1}^{N/2} |\pi_i \rangle \langle \pi_i|\right),
\end{equation}
expressed in the position basis, which implies a von Neumann entropy of $S_{\mathrm{vN}} = \ln(2)$. Away from the critical point the chain is gapped and the bipartite entanglement entropy obeys an area law, saturating with subsystem size. 
At the critical point, however, the gap closes. The entanglement entropy then grows logarithmically with subsystem size, $S_{\mathrm{vN}} \sim (c/6)\ln N$. The central charge is $c = 1/2$ for the $(1+1)d$ Ising universality class.\cite{calabrese2004entanglement,eisert2010area,serwatka2024quantum}
This logarithmic peak makes the entanglement entropy a powerful probe of the phase diagram of the rotor chain. This is relevant both in the presence and in the absence of the six-fold potential.

\section{Results and Discussion\label{paper2:results}}

\subsection{Monomer}
We first solve the monomer problem to establish the basis size required to converge the first few eigenstates and determine where they lie with respect to the reduced barrier height $A$. 
Neutron diffraction difference-Fourier maps of the proton density in beryl reveal a six-fold-modulated, delocalized distribution.\cite{kolesnikov2016quantum} This observation motivates a quantitative analysis of the monomer levels in a six-fold potential.

We compute the eigenvalues of the monomer Hamiltonian of Eq.~\eqref{paper2:eq:h1B} for barrier heights up to $A=80$, the largest value considered in the present work.
We report the error in the tunneling splitting, $\Delta=E_1-E_0$, for $m_{\mathrm{max}}=16$ in Table~\ref{paper2:tab:gapconv}.

\begin{table}[htbp]
\caption{\label{paper2:tab:gapconv}Convergence of the single-rotor tunneling splitting
$\Delta=E_1-E_0$ with respect to the size of the angular-momentum basis. Values
obtained at $m_{\max}=16$ are compared with converged results
($m_{\max}=90$).}
\begin{ruledtabular}
\begin{tabular}{ccccc}
$A$ & $\Delta$ (converged) & $\Delta$ ($m_{\max}=16$) &
$|\delta\Delta|$ & Error (\%) \\
\hline
$20$ & 0.8498913 & 0.8498941 & $2.86\times10^{-6}$ & 0.00034 \\
$40$ & 0.5728426 & 0.5729498 & $1.07\times10^{-4}$ & 0.019 \\
$60$ & 0.3525340 & 0.3531961 & $6.62\times10^{-4}$ & 0.188 \\
$80$ & 0.2134483 & 0.2155106 & $2.06\times10^{-3}$ & 0.966 \\
\end{tabular}
\end{ruledtabular}
\end{table}

\begin{figure}
    \centering
    \includegraphics[width=\linewidth]{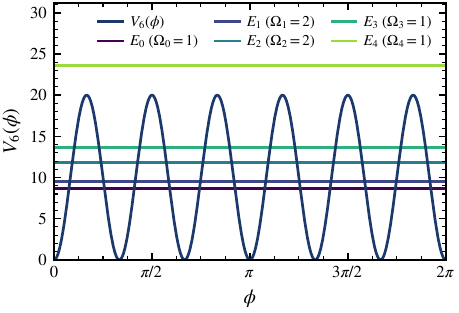}
    \caption{Energy levels of a single rotor in a six-fold potential with the barrier height $A=20$; $\Omega_i$ denotes the degeneracy of each level. Based on the convergence tests, $m_{\mathrm{max}}=16$ is used.}

    \label{paper2:fig:mono_state_A20}
\end{figure}

The distribution of energy levels (and their degeneracies) in the six-fold potential is illustrated in Fig.~\ref{paper2:fig:mono_state_A20} for $A=20$.
With this value, six states have an energy below the barrier.
\begin{figure}
    \centering
    \includegraphics[width=\linewidth]{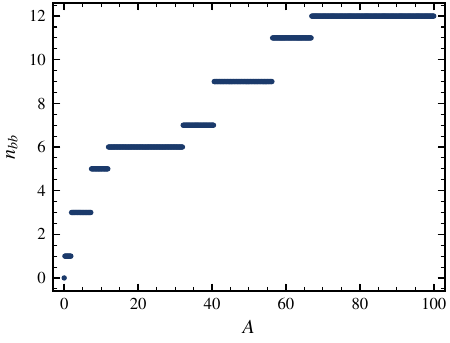}
    \caption{Number of states below the barrier ($n_{bb}$) as a function of the six-fold (reduced) barrier height $A$.}

    \label{paper2:fig:numstates_V6}
\end{figure}
As depicted in Fig.~\ref{paper2:fig:numstates_V6}, $n_{bb}$ reaches 6 when $A \approx 18$ and stays on this first plateau until $A \approx 38$. Upon further increase to $A \approx 70$, $n_{bb}$ climbs to 12, forming a second plateau. Each plateau corresponds to a filled manifold of six (nearly degenerate) well-localized states, one per minimum of the six-fold potential.
When the six-fold potential is relatively small, as shown in Fig.~\ref{paper2:fig:numstates_V6}, only the ground state lies below the barrier of the cosine potential. To enable the dipole rotor to tunnel from the current orientation to another local minimum, it is sufficient to excite the dipole to the first excited state. Increasing $A$ brings more excited states below the barrier. 
Model calculations accompanying the INS experiments predict tunneling transitions at 2.7, 10.1 and 16.7 meV, the first two doubly degenerate.\cite{kolesnikov2016quantum}
Thus, reorientation between wells proceeds either by tunneling within the below-barrier manifold or by kinetic excitation above the barrier. The number of below-barrier states controls which channel dominates.

\subsection{Longer chains: $N=50$}
In order to optimize the MPS representation of the ground states, we employ a two-site DMRG algorithm as implemented in ITensor.\cite{fishman2022itensor} An SVD threshold of $10^{-8}$ and a relative energy cut-off of $10^{-8}$ are used. Local Hilbert spaces are described by a plane wave eigenbasis with $m_{\mathrm{max}}=16$. This basis size was established by converging the monomer eigenstates in the preceding subsection.
The maximum bond dimension retained during the sweeps was $200$, and $100$ sweeps were sufficient to converge the ground-state energy to the stated tolerance. We also verified for representative points at the largest $A$ and $g$ that increasing the momentum cutoff beyond $m_{\mathrm{max}}=16$ leaves the reported observables unchanged. Deep in the ordered phase the ground state becomes nearly two-fold degenerate. The sweeps reported here converge to the symmetric superposition of Eq.~\eqref{paper2:gs_order}. This is consistent with the observed plateaus $U_x \to 2/3$ and $S_{\mathrm{vN}} \to \ln 2$. The moments $\braket{M_x^2}$ and $\braket{M_x^4}$ were evaluated directly from the converged MPS as expectation values of the corresponding operator sums.

\begin{figure}
    \centering
    \includegraphics[width=\linewidth]{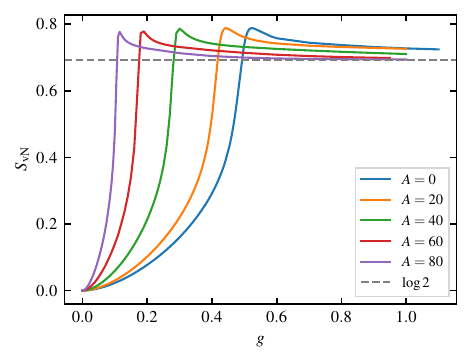}
    \caption{Von Neumann entropy $S_{\mathrm{vN}}$ as a function of $g$ for $A=0,20,40,60,80$, and $N=50$.}

    \label{paper2:fig:svn}
\end{figure}

Fig.~\ref{paper2:fig:svn} shows the von Neumann entropy $S_{\mathrm{vN}}$ as a function of the dipolar coupling $g$ for several reduced barrier heights $A$ at $N=50$. As $A$ increases, the transition shifts to smaller $g$ compared with free rotors ($A=0$), for which $g_c \approx 0.5$.\cite{serwatka2024quantum,iouchtchenko2018ground} In particular, for larger $A$ the entropy reaches its plateau at smaller $g$. This demonstrates the tuning capability offered by hindered dipolar rotors.

\begin{figure}
    \centering
    \includegraphics[width=\linewidth]{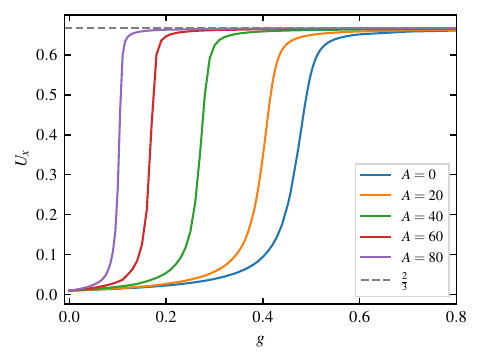}
    \caption{Binder cumulant $U_{x}$ as a function of the dipolar coupling $g$ for $A=0,20,40,60,80$ and $N=50$.}

    \label{paper2:fig:U_x}
\end{figure}

\begin{figure}
    \centering
        \includegraphics[width=\linewidth]{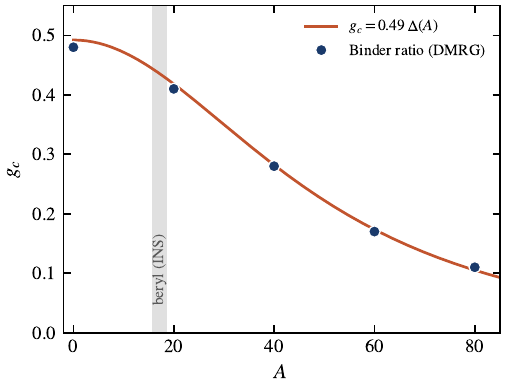}
\caption{\label{paper2:fig:phase_diagram}Critical coupling $g_c$ as a function of the
reduced barrier height $A$. Symbols: $g_c$ from the maximum gradient of the
Binder ratio at $N=50$ (Fig.~\ref{paper2:fig:U_x}). Solid line: the scaling form
$g_c = a\,\Delta(A)$ of Eq.~\eqref{paper2:eq:gc_scaling} with $a = 0.49$, where
$\Delta(A) = E_1 - E_0$ is the tunneling splitting of the monomer
Hamiltonian of Eq.~\eqref{paper2:eq:h1B}. The $A=0$ point is not included in the
fit; the form nevertheless recovers the free-rotor value $g_c \approx 1/2$
there, since $\Delta(0)=1$. The shaded band marks the beryl barrier range
$A \approx 16$--$18$ from inelastic neutron
scattering,\cite{kolesnikov2016quantum} where the fit gives
$g_c \approx 0.44$.}
\end{figure}

The Binder ratio $U_x$ (Fig.~\ref{paper2:fig:U_x}) is computed for the same barrier heights ($A = 0, 20, 40, 60, 80$) and the same DMRG parameters. It displays the analogous trend.
The plateau at $U_x = 2/3$ is reached at smaller $g$ as $A$ increases. The critical couplings $g_c(A)$, extracted from the maximum gradient of the Binder ratio, are collected in the phase diagram of Fig.~\ref{paper2:fig:phase_diagram}.

\begin{figure}[htbp]
    \centering
    \includegraphics[width=\linewidth]{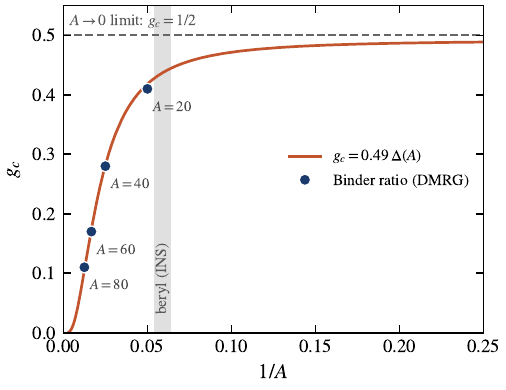}
    \caption{\label{paper2:fig:gc_scaling}Critical coupling $g_c$ as a function of
    the inverse reduced barrier height $1/A$. Symbols: $g_c$ extracted from
    the maximum gradient of the Binder ratio at $N=50$
    (Fig.~\ref{paper2:fig:U_x}). Solid line: the single-parameter form
    $g_c = a\,\Delta(A)$ with $a = 0.49$, where
    $\Delta(A) = E_1 - E_0$ is the tunneling splitting of the monomer
    Hamiltonian of Eq.~\eqref{paper2:eq:h1B}. The line extrapolates the phase
    boundary at both ends: for $A \to \infty$ ($1/A \to 0$) the splitting,
    and with it $g_c$, vanishes exponentially, while for $A \to 0$
    ($1/A \to \infty$) it approaches the free-rotor limit $g_c = 1/2$
    (dashed line), since $\Delta(0) = 1$ exactly. The shaded band marks the
    barrier range $A \approx 16$--$18$ inferred for water in beryl from
    inelastic neutron scattering,\cite{kolesnikov2016quantum} for which the
    scaling form gives $g_c \approx 0.43$.}
\end{figure}

The dependence of the critical coupling on the barrier height is captured by
a simple scaling relation. For each barrier height, $g_c$ is extracted from
the maximum gradient of the Binder ratio $U_x(g)$ at $N=50$
(Fig.~\ref{paper2:fig:U_x}).
Independently, the monomer tunneling splitting $\Delta(A) = E_1 - E_0$ is
obtained by exact diagonalization of the monomer Hamiltonian of
Eq.~\eqref{paper2:eq:h1B}. A converged momentum cutoff is used
($m_{\max} = 90$, cf.\ Table~\ref{paper2:tab:gapconv}).
The ratio
$g_c/\Delta$ is found to be constant over the entire range of barriers
studied, so that
\begin{equation}
    g_c(A) \simeq a\,\Delta(A), \qquad a = 0.49,
    \label{paper2:eq:gc_scaling}
\end{equation}
where $a$ is the mean of $g_c/\Delta$ over $A = 20$--$80$. This form
interpolates the computed points and fixes both asymptotes. 
In the free-rotor limit $\Delta(0) = 1$ one recovers $g_c = 1/2$.
For large barriers $\Delta$ decays exponentially and the ordered phase extends to
arbitrarily weak coupling. 
The relation is analogous to the transverse-field-Ising case.
The dipole operator couples the monomer ground state to the first excited level (the dipole-active member of the lowest sextet). At low energies, each hindered rotor therefore acts as a two-level system with gap $\Delta(A)$. Criticality occurs when the dipolar coupling balances this quantum-fluctuation scale at $g_c \simeq \Delta(A)/2$.
The solid line in
Fig.~\ref{paper2:fig:gc_scaling} is Eq.~\eqref{paper2:eq:gc_scaling} evaluated on a dense
grid of barrier heights. At the beryl-relevant barrier, it yields
$g_c \approx 0.43$, well below the estimated physical coupling
$g \approx 0.90$.

\begin{table*}[htbp]
\caption{\label{paper2:tab:barriers}Reported values of the in-plane orientational barrier for type-I water in beryl. The different methods extract different quantities (see text). For orientation, 1~meV $\approx 8.07$~cm$^{-1}$ $\approx 11.6$~K.}
\begin{ruledtabular}
\begin{tabular}{llc}
Method & Quantity extracted & Value (meV) \\
\hline
INS (tunneling)\cite{kolesnikov2016quantum} & barrier between minima (static) & $\approx 176$ \\
INS (tunneling)\cite{kolesnikov2016quantum} & barrier between minima (vibration-lowered) & $48$--$56$ \\
Two-level model\cite{klivc2023orientational} & six-fold barrier $V_b$ (bounds adopted from Ref.~\onlinecite{kolesnikov2016quantum}) & $56$--$176$ \\
THz spectroscopy\cite{gorshunov2016incipient} & six-well depth (fit) & $1.41 \pm 0.05$ \\
DFT\cite{gorshunov2016incipient} & six-well depth (DFT) & $\approx 0.83$\\
Dielectric relaxation\cite{anovitz2013anisotropic} & reorientation activation energy & $170 \pm 1.5$ \\
\end{tabular}
\end{ruledtabular}
\end{table*}

\subsection{Connection to beryl parameters}

Earlier studies on the vibrational energy levels of confined water molecules in beryl demonstrated that these water molecules form virtually negligible hydrogen bonds with the cage wall.\cite{kolesov2000orientation} Subsequently, Raman spectroscopic analysis indicated that type-I water molecules exhibit a certain degree of motion at a frequency that can be approximated by the ``particle-in-a-box'' model. The corresponding frequencies suggest that the water molecules undergo anisotropic rotation about at least two of the three possible rotational axes within the cavity.\cite{kolesov2008vibrational} Inelastic neutron scattering (INS) spectra were analyzed to extract the energies and intensities of the excitations. In contrast to water in C$_{60}$,\cite{felker2017explaining,suzuki2019rotational} the absence of transitions between the para and the ortho states implies the water was trapped in a way that the dipole is perpendicular to the c-axis.\cite{kolesnikov2016quantum} 
Such planar confinement and the multi-well potential energy landscape are further supported by recent investigations on structurally related cordierite lattices. These studies revealed rich single-particle tunneling and collective dipolar excitations within the basal plane.\cite{belyanchikov2022single} The beryl nano-cage thus provides planar confinement with intrinsic six-fold symmetry. A chain of planar rotors with a six-fold ($\mathbb{Z}_6$) on-site potential is therefore a promising framework to describe the tunneling dynamics of the confined water molecules.

Beryl can accommodate coplanar chains with a distance of $R = 9.2$~\AA{} between the planar rotors.\cite{kolesnikov2016quantum,gorshunov2016incipient,anovitz2013anisotropic,klivc2023orientational}
The distance of 9.2~\AA{} corresponds to nearest cages in \emph{adjacent} channels, lying in a common plane perpendicular to the crystal $c$-axis. It is this in-plane arrangement that realizes the coplanar chain of Fig.~\ref{paper2:fig:rotor_chain_with_potential}.\cite{gorshunov2016incipient} 
Cages within a single channel are closer (4.6~\AA). 
For that stacking direction, however, the water dipoles are perpendicular to the inter-rotor separation.
In that case, antiparallel alignment is favored and is not described by the coplanar chain model considered here.
The chain studied in this work is therefore an idealized subsystem of the full three-dimensional dipolar lattice of beryl.
Using a value of $\mu=1.85$ D for the dipole moment of water along with a rotational constant of $B=3.04$ meV,\cite{kolesnikov2016quantum} we obtain the following value for the interaction strength,
\begin{equation}
    g=\frac{|\mu|^2}{B ~4\pi\epsilon_0 R^3}=0.90.
\end{equation}

The value of the six-fold barrier itself is less settled, as the available estimates span two orders of magnitude. Table~\ref{paper2:tab:barriers} collects the values reported in the literature. 
The spread largely reflects the fact that the different techniques extract \emph{different} quantities, all loosely referred to as \emph{the barrier}.
The INS tunneling analysis yields a static barrier between minima of the orientational potential of $\approx 176$~meV.
That quantity is lowered to 48--56~meV
($\approx 390$--$450$~cm$^{-1}$, $\approx 560$--$650$~K) when the coupling to cage vibrations is taken into account.\cite{kolesnikov2016quantum} 
The lower and upper bounds of 56 and 176~meV used in a subsequent theoretical study\cite{klivc2023orientational} are not an independent measurement. 
They are obtained directly from that INS analysis where the upper bound is the static barrier and the lower bound is the vibration-lowered value. 
In that study the potential is written in a coupled two-level form rather than as a plain $\cos(6\phi)$ term. Conversely, the much smaller THz-derived value of $1.41 \pm 0.05$~meV (with $\approx 0.83$~meV from DFT)\cite{gorshunov2016incipient} is the depth parameter of the six-well surface probed by the dipole-active transitions. This is a different quantity from the barrier through which the proton density tunnels. Broadband dielectric relaxation measurements\cite{anovitz2013anisotropic} yield an Arrhenius activation energy for dipole reorientation of $16.4 \pm 0.14$~kJ/mol ($\approx 170$~meV). Again, this is a kinetic rather than a static quantity. Its value is quite close to the static INS barrier.\cite{kolesnikov2016quantum} 
Reconciling these observables within a single six-fold potential remains an open question.

Here, we adopt the INS-based barrier range of 48--56~meV.\cite{kolesnikov2016quantum,klivc2023orientational} The corresponding (reduced) dimensionless barrier is $A = A'/B \simeq 48/3.04$ to $56/3.04 \approx 16$--$18$. We can extract the critical coupling from the phase diagram shown in Fig.~\ref{paper2:fig:phase_diagram} at this barrier value. We obtain $g_c \simeq 0.4$. 
Since the physical coupling $g \approx 0.90$ exceeds $g_c$, this estimated value places coplanar chains of water in beryl in the ordered (ferroelectric) phase.
We note that this conclusion is robust against the uncertainty in the barrier height. Even in the nearly-free-rotor limit suggested by the THz analysis ($A \approx 1.41/3.04 \approx 0.5$), the critical coupling is $g_c \approx 0.5$. The estimated $g \approx 0.90$ still exceeds that number. Barriers larger than the INS value only shift $g_c$ to smaller values. For instance, $A \approx 176/3.04 \approx 58$ for the upper bound of Ref.~\onlinecite{klivc2023orientational}, a value well within the range of Fig.~\ref{paper2:fig:phase_diagram}. Coplanar water chains in beryl are therefore predicted to lie in the ordered phase for the entire reported range of barrier heights.

\section{Concluding Remarks\label{paper2:conclusions}}
We have studied the energetic, structural, and entanglement properties of a chain of hindered planar rotors with a six-fold potential of reduced barrier height $A$. The monomer energy levels were first computed for $A=20$. The number of states below the barrier ($n_{bb}$) exhibits a regular staircase structure as a function of the reduced barrier height $A$. The rotor chain composed of these planar rotors undergoes a quantum phase transition from the disordered to the ordered phase as the dipolar coupling increases. 
The transition was characterized using the von Neumann entropy and the Binder cumulant. It reflects the competition between the dipolar interaction and the single-rotor kinetic energy. A clear trend emerges: as $A$ increases, the von Neumann entropy of Fig.~\ref{paper2:fig:svn} reaches its plateau near $\ln 2$ at weaker interaction strength $g$.
The Binder cumulant $U_x$, shown in Fig.~\ref{paper2:fig:U_x}, exhibits a similar trend. Indeed, $U_x$ approaches its plateau of $2/3$ at smaller $g$ as $A$ increases.  
Previous works using a number of these techniques suggest that the ground state of a dipolar planar rotor chain admits an ordered phase as well as a disordered phase.\cite{serwatka2024quantum} 
Both phases can be accessed by tuning the dipolar coupling strength. 
Increasing the reduced barrier height from $A=0$ to $A=80$ lowers the critical coupling from $g_c \approx 0.5$ to substantially smaller values.
A central result of this work is the collapse of the hindered-chain phase boundary onto the monomer tunneling splitting, $g_c \simeq \Delta(A)/2$. This relation is simple and physically transparent, and it should prove useful beyond the beryl application considered here.
Our parameter estimates place a coplanar chain of water molecules in beryl in the ordered (ferroelectric) phase. Whether the idealized coplanar chains can be found in beryl samples remains an open question.

Experimentally, however, beryl shows only incipient ferroelectricity, with no long-range order down to sub-kelvin temperatures.\cite{gorshunov2016incipient,belyanchikov2020dielectric} The absence of order may reflect partial cage occupancy or frustrating inter-chain couplings.
The current results reveal a crossover behavior at finite $N$. Future work will include the required finite-size-scaling studies in order to obtain critical exponents.
We also plan to perform finite-temperature path integral simulations for higher-dimensional lattices.\cite{zhang2025path,Moeed2025,vb2025path}

\section*{Acknowledgements}
The authors acknowledge support from NSERC (RGPIN-03725-2022), the Ontario Ministry of Research and Innovation (MRI), the Canada Research Chair program (950-231024), the Digital Research Alliance of Canada, and the Canada Foundation for Innovation (CFI) (project No. 35232).

\section*{DATA AVAILABILITY}
The data that support the findings of this study are available from the corresponding author upon reasonable request.

\bibliography{literature}

\end{document}